\documentstyle[12pt]{article}

\newcommand{\be}{\begin{equation}}
\newcommand{\ee}{\end{equation}}
\newcommand{\ks}{k\!\!\!/}
\newcommand{\as}{a\!\!\!/}

\date{}
\title{
{\large\rm DAMTP-97-10}\hfill\vspace*{0cm}\\
{\large\rm February 1997}\hfill\vspace*{2.5cm}\\
Diffractive Parton Distributions\\ in the Semiclassical Approach
\\[2cm]}
\author{
A. Hebecker\\
{\normalsize\it D.A.M.T.P., Cambridge University, Cambridge CB3 9EW, England}
\vspace*{3cm}\\                     
}                                                                          
\input psfig
\begin{document}

\setlength{\baselineskip}{18pt}
\maketitle
\begin{abstract}
\noindent
Recently, a semiclassical approach to diffraction has been proposed, which 
treats the proton as a classical colour field. The present paper 
demonstrates that this approach is consistent with the concept of 
diffractive parton distributions. The diffractive quark and gluon 
distributions are expressed through integrals of non-Abelian eikonal 
factors in the fundamental and adjoint representation respectively. As a 
by-product, previously calculated diffractive cross sections for processes 
with a final state gluon are rederived in a simpler way. 
\end{abstract}
\thispagestyle{empty}
\newpage

\section{Introduction}
The measurement of diffractive structure functions in deep inelastic 
scattering at HERA \cite{ex} has triggered renewed theoretical interest in 
the phenomenon of hard diffraction (which was seen previously in 
hadron-hadron collisions \cite{ua8}). A particularly important observation 
is the absence of a large-$Q^2$ suppression. This suggests a significant 
leading twist component in diffraction, which is the main subject of the 
present analysis. 

One of the interesting theoretical problems is the description of the 
colour singlet exchange, responsible for the large rapidity gap in the 
final state. A large amount of theoretical work has been devoted to two 
gluon exchange models (see e.g. \cite{tg}), which correspond to the 
simplest possible perturbative description. It is, however, not obvious 
that such a perturbative approach is valid for the bulk of the events. 

A different approach is based on the picture of soft pomeron exchange. 
The hardness of the process, introduced by the large scale $Q^2$, suggests 
a partonic interpretation of the pomeron \cite{is}. As pointed out by 
Berera and Soper \cite{bs1}, instead of applying the parton model to the 
pomeron, one can also introduce the more general but less predictive 
concept of diffractive parton distributions, closely related to the 
fracture functions of \cite{vt}. This approach, and the underlying concept 
of diffractive factorization, has been discussed in more detail in 
\cite{bs}. 

The semiclassical approach derives leading twist diffraction from soft 
interactions of fast partons with the proton colour field \cite{bh,bdh}. 
Within this model the dominant contribution to diffractive $q\bar{q}$-pair 
production comes from aligned jet type configurations, which interact with 
the target in a soft process. Therefore, the semiclassical model is similar 
in spirit to the non-perturbative models discussed above, although it does 
not use the concept of a pomeron. 

It has been suggested that diffractive factorization can be understood 
in the semiclassical picture in the proton rest frame \cite{soper}. The 
present paper demonstrates the consistency of the semiclassical calculation 
with the concept of diffractive structure functions. It is explicitly shown 
that the amplitude contains two fundamental parts: the usual hard 
scattering amplitude of a partonic process, and the amplitude for soft 
eikonal interactions with the external colour field. The latter part is 
dominated by the scattering of one of the partons from the photon wave 
function, which has to have small transverse momentum and to carry a 
relatively small fraction of the photon energy in the proton rest frame. In 
a frame where the proton is fast this parton can be interpreted as a 
parton from the pomeron structure function. 

Calculating the cross section by standard methods, a result is obtained 
that can be written as a convolution of a partonic cross section and a 
diffractive parton distribution. Within the semiclassical model this 
diffractive parton distribution is explicitly given in terms of integrals 
of non-Abelian eikonal factors in the background field. 

The special r\^{o}le played by the soft parton in the photon wave function 
has also been discussed in \cite{nz} in the framework of two gluon 
exchange. However, the present approach has the two following advantages: 
firstly, by identifying the hard part as a standard photon-parton 
scattering cross section the necessity for non-covariant photon wave 
function calculations is removed. Secondly, once it is established that the 
main contribution comes from the soft region, non-perturbative effects are 
expected to become important. The eikonal approximation provides a simple, 
self-consistent model for this non-perturbative region. 

The paper is organized as follows. In Sect.~\ref{scal} the semiclassical 
calculation of the diffractive cross section is performed for the simple 
scalar case. The corresponding diffractive distribution of scalar partons 
is derived. Sections~\ref{qua} and \ref{glu} introduce the appropriate 
changes to obtain the diffractive quark and gluon distributions 
respectively. The application of the developed formalism to the simplest 
hard diffractive processes of $q\bar{q}$ and $q\bar{q}g$ state production 
is discussed in Sect.~\ref{app}. In particular, the agreement with the 
results of \cite{bdh} is demonstrated. The conclusions in Sect.~\ref{con} 
are followed by two appendices explaining some technical details relevant 
to Sections~\ref{qua} and \ref{glu}.

\section{Diffractive distribution of scalar partons}\label{scal}
In this section a simple, yet technically sufficiently interesting, model 
process will be investigated: the incoming virtual photon with momentum $q$ 
($q^2=-Q^2,\, x\ll 1$) is assumed to fluctuate into a set of scalar 
coloured partons. This virtual state is then made real by its interaction 
with the proton, resulting in a diffractive final state with invariant mass 
$M$ and an elastically scattered proton. 

Following the concept introduced in \cite{bh} and further developed in 
\cite{bdh}, the proton is treated as a soft external colour field. In the 
high energy limit the only effect of this field is to introduce a 
non-Abelian phase factor for each of the coloured particles passing the 
field. 

\begin{figure}[ht]
\begin{center}
\vspace*{-.5cm}
\parbox[b]{10cm}{\psfig{width=9cm,file=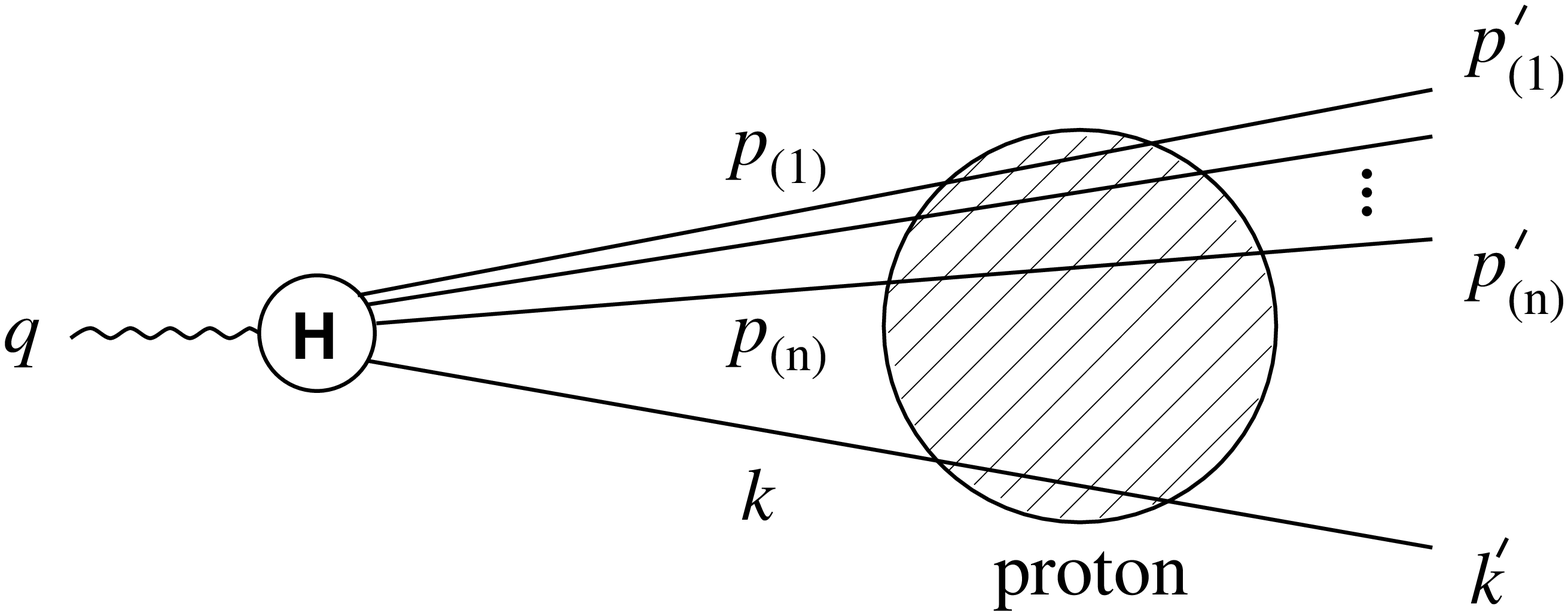}}\\
\end{center}
\refstepcounter{figure}
\label{process}
{\bf Fig.\ref{process}} Hard diffractive process in the proton rest frame. 
The soft parton with momentum $k$ is responsible for the leading twist 
behaviour of the cross section. 
\end{figure}

The kinematical situation is shown symbolically in Fig.~\ref{process}. The 
process is split into two parts, the hard amplitude for the transition of 
the photon into a virtual partonic state and the scattering of this state 
off the external field. 

To keep the amplitude for the first part (symbolized by a circled ``H'' in 
Fig.~\ref{process}) hard the transverse momenta $p_{(j)\perp}'$ 
$(j=1...n)$ are required to be large, i.e. of order $Q$. The momentum 
$k_\perp$ is small, i.e. of the order of some hadronic scale $\Lambda$, and 
the corresponding parton carries only a small fraction ($\sim\Lambda^2/ 
Q^2$) of the longitudinal photon momentum in the proton rest frame. While 
the hardness condition for particles 1 through $n$ is introduced 
``by hand'', simply to make the process tractable, the softness of the last 
particle follows automatically from the requirement of leading twist 
diffraction. This will become clear from the calculation below (see also 
\cite{bdh}). 

The rest of this section is devoted to the calculation of the above 
diffractive process using the eikonal model for the scattering off the 
proton field. The result will be a convolution of a hard photon-parton 
scattering cross section, as it is typical for deep inelastic scattering 
seen, e.g., in the Breit frame (brick wall frame), with a diffractive 
parton distribution calculated from the soft interaction of a high energy 
parton and the external colour field. 

The standard cross section formula for the scattering off a static external 
field reads
\be
d\sigma=\frac{1}{2q_0}\,|T|^2\,2\pi\delta(q_0-q_0')\,dX^{(n+1)}
\,,\quad\mbox{where}\quad q'=k'+\mbox{$\sum$}p_{(j)}'\, .\label{cs1}
\ee
All momenta are given in the proton rest frame, $T$ is the amplitude 
corresponding to Fig.~\ref{process}, and $dX^{(n+1)}$ is the usual phase 
space element for $n+1$ particles.

Each of the particles scatters off the external field with an effective 
vertex \cite{css}
\be
V(p',p)=2\pi\delta(p_0'-p_0)\,2p_0\,\tilde{U}(p_\perp'-p_\perp)
\,,\label{hev}
\ee
where $\tilde{U}$ is the Fourier transform of the non-Abelian eikonal 
factor in impact parameter space,
\be
U(x_\perp)=P\exp\left(-\frac{i}{2}\int_{-\infty}^{\infty}A_-(x_+,x_\perp)
dx_+\right)\, .
\ee
Here $x_\pm=x_0\pm x_3$ are the light-cone components of $x$, 
$A(x_+,x_\perp)$ is the gauge field, and the path ordering $P$ sets the 
field at smallest $x_+$ to the rightmost position. 

This results in the following equation for the combined amplitude $T'$, 
\[
i\,2\pi\delta(q_0-q_0')\,T'=\int T_H\,\prod_j\!\left(\frac{i}{p_{(j)}^2}
\,2\pi\delta(p_{(j)0}'-p_{(j)0})\,2p_{(j)0}\,\tilde{U}(p_{(j)\perp}'-
p_{(j)\perp})\,\frac{d^4p_{(j)}}{(2\pi)^4}\right) 
\]
\be
\hspace*{5cm}\times\left(\frac{i}{k^2}\,2\pi\delta(k_0'-k_0)\,2k_0\,
\tilde{U}(k_\perp'-k_\perp)\right)\, ,\label{amp1}
\ee
where $T_H$ stands for the part of the diagram in Fig.~\ref{process} that 
is symbolized by the circled ``H''. The prime on the amplitude shows that 
the above expression is not yet exactly the amplitude required for 
Eq.~(\ref{cs1}). The r.h.s. of Eq.~(\ref{amp1}) includes the unphysical 
contribution where none of the partons interacts with the field, 
corresponding to the zeroth order term in an expansion of all $U$'s in 
powers of $A$. It is convenient to subtract this trivial term later on by 
hand. 

Colour indices have been suppressed in Eq.~(\ref{amp1}). Notice also, that 
some of the produced partons are antiparticles. The corresponding matrices 
$U$ have to be replaced by $U^\dagger$. To keep the notation simple, this 
has not been indicated explicitly. 

The integrations over the light-cone components $p_{(j)+}$ can be performed 
using the appropriate energy $\delta$-functions. After that, the 
$p_{(j)-}$-integrations are performed by picking up the poles of the 
propagators $1/p_{(j)}^2$. The result is 
\be
T=\int T_H\,\prod_j\!\left(\tilde{U}(p_{(j)\perp}'-p_{(j)\perp})\,\frac{d^2
p_{(j)\perp}}{(2\pi)^2}\right) \frac{2k_0}{k^2}\,\tilde{U}(k_\perp'-k_\perp)
\, .\label{amp2}
\ee

Next, a change of integration variables is performed, 
\be
d^2p_{(n)\perp}\,\to\, d^2k_\perp\,.
\ee
Since the external field is assumed to be soft, it can only transfer 
transverse momenta of order $\Lambda$, i.e., $p_{(j)\perp}\approx 
p_{(j)\perp}'$ for all $j$. In general, the amplitude $T_H$ will be 
dominated by the hard momenta of order $Q$. Therefore, it can be assumed 
that $T_H$ is constant if the momenta $p_{(j)\perp}$ vary on a scale 
$\Lambda$. In this approximation the integrations over $p_{(j)\perp}$ 
$(j=1...n-1)$ can be performed in Eq.~(\ref{amp2}), resulting in 
$\delta$-functions in impact parameter space. These manipulations give 
the result 
\be
T'=\int\frac{d^2k_\perp}{(2\pi)^2}\,\frac{2k_0}{k^2}\,T_H
\int_{x_\perp,y_\perp}\bigg(\prod_jU(x_\perp)\bigg)U(x_\perp+y_\perp)
\,e^{-ix_\perp\Delta_\perp-iy_\perp(k_\perp'-k_\perp)}\,\, ,\label{amp3}
\ee
where $\Delta$ is the total momentum transferred from the proton to the 
diffractive system and, in particular, $\Delta_\perp=k_\perp'+\sum 
p_{(j)\perp}'$. It is intuitively clear that the relative proximity of the 
high-$p_\perp$ partons in impact parameter space leads to the corresponding 
eikonal factors being evaluated at the same position $x_\perp$. 

Now the colour structure of the amplitude will be considered in more 
detail. Spelling out all the colour indices and introducing explicitly the 
colour singlet projector $P$ the relevant part of the amplitude reads,
\be
T_{colour}'=T_H^{a_1...a_nb}\,\bigg(\prod_jU(x_\perp)
\bigg)^{a_1'...a_n'}_{a_1...a_n}\,U(x_\perp+
y_\perp)^{b'}_b\,P_{a_1'...a_n'b'}\, .
\ee
Using the fact that $T_H$ is an invariant tensor in colour space and 
introducing the function
\be
W_{x_\perp}(y_\perp)^{b'}_b=\left(U(x_\perp)^\dagger U(x_\perp+y_\perp)-1
\right)^{b'}_b\, ,\label{wdef}
\ee
the following formula is obtained,
\be
T_{colour}=T_{colour}'-T_H^{a_1...a_nb}P_{a_1...a_nb}=
T_H^{a_1...a_nb}\,W^{b'}_b P_{a_1...a_nb'}\, .\label{tcol}
\ee
Here the first equality has to be understood as the definition of the 
corrected amplitude, where the trivial contribution of zeroth order in $A$ 
has been subtracted. This subtraction has been taken into account in the 
definition of $W$, Eq.~(\ref{wdef}). 

For colour covariance reasons
\be
T_H^{a_1...a_nb} P_{a_1...a_nb'}=\mbox{const.}\,\times\,\delta^b_{b'}\, .
\label{const}
\ee
Since the photon is colour neutral, the following equality holds,
\be
T_H^{a_1...a_nb}\,T^*_{H\,a_1...a_nb}=|\,T_H^{a_1...a_nb} P_{a_1...a_nb}\,
|^2=|\mbox{const.}|^2N^2\, .\label{const1}
\ee
Here the partons are assumed to be in the fundamental representation of 
the colour group $SU(N)$. Combining Eq.~(\ref{tcol}) with 
Eqs.~(\ref{const}) and (\ref{const1}) it becomes clear that the colour 
structure of the hard part decouples from the eikonal factors, 
\be
|T_{colour}|^2=\frac{1}{N}\,|\,\mbox{tr}[W]\,|^2\,\,|T_H|^2\, .\label{tc}
\ee
The hard part will be interpreted in terms of an incoming small-$k_\perp$ 
parton, which collides with the $\gamma^*$ to produce the outgoing partons 
1 through $n$. Therefore a factor $1/N$ for initial state colour 
averaging has been included into the definition of $|T_H|^2$. 

In the expression for the cross section the two functions $W$ enter in the 
combination 
\be
\Big(\mbox{tr}[W_{x_\perp}(y_\perp)]\Big)\Big(\mbox{tr}[W_{x_\perp'}
(y_\perp')]\Big)^*\,e^{-i(x_\perp-x_\perp')\Delta_\perp}\, ,\label{ww}
\ee
with independent integrations over $x_\perp,x_\perp',y_\perp$ and 
$y_\perp'$. If the external field is sufficiently smooth, the functions $W$ 
vary only slowly with $x_\perp$ and $x_\perp'$. Therefore, after 
integration over $x_\perp$ and $x_\perp'$ the expression in Eq.~(\ref{ww}) 
will produce an approximate $\delta$-function in $\Delta_\perp$. 
Furthermore, it will be assumed that the measurement is sufficiently 
inclusive, i.e., the hard momenta $p_{(j)_\perp}'$ are not resolved on a 
soft scale $\Lambda$. This corresponds to a $\Delta_\perp$-integration, 
which will produce an approximate $\delta$-function in $x_\perp\!- 
x_\perp'$. Since the expression in Eq.~(\ref{ww}) will always appear under 
$x_\perp,\,x_\perp'$- and $\Delta_\perp$-integration, the above 
considerations justify the substitution 
\be
e^{-i(x_\perp-x_\perp')\Delta}\to (2\pi)^2\,\delta^2(x_\perp-x_\perp')\,
\delta^2(\Delta_\perp)\, .
\ee
Combining Eqs.~(\ref{cs1}),(\ref{amp3}), and (\ref{tc}) the following 
formula for the cross section results, 
\be
d\sigma\!=\frac{1}{2q_0}\int |T_H|^2 \int_{x_\perp}\left|\int\frac{d^2
k_\perp}{(2\pi)^2}\,\frac{\mbox{tr}[\tilde{W}_{x_\perp}(k_\perp'\!\!\!-\!
k_\perp)]}{\sqrt{N}\,k^2}\right|^2\!\!(2k_0)^2(2\pi)^3\,\delta^2\!\left(
\mbox{$\sum$}p_{(j)\perp}\right)\delta(q_0\!\!-\!q_0')\,dX^{(n+1)}.
\label{cs2}
\ee
Note that the soft momentum $k_\perp'$ has been neglected in the transverse 
$\delta$-function. 

To finally establish the parton model interpretation of diffraction the 
hard partonic cross section based on $|T_H|^2$ has to be identified in 
Eq.~(\ref{cs2}). Consider the process
\be
\gamma^*(q)+q(yP)\to q(p_{(1)}')+\dots+q(p_{(n)}')\, ,\label{pp}
\ee
where the photon collides with a parton carrying a fraction $y$ of the 
proton momentum and produces $n$ high-$p_\perp$ final state partons. The 
cross section is approximately given by
\be
d\hat{\sigma}(y)=\frac{1}{2(\hat{s}+Q^2)}\,|T_H|^2\,(2\pi)^4\delta^4(q-k-
\mbox{$\sum$}p_{(j)}')\,dX^{(n)}\, ,\label{pcs}
\ee
where $\hat{s}=(\sum p_{(j)}')^2$ and the quantities $|T_H|^2$ and $k$ are 
the same as in the previous discussion. Eq.~(\ref{pcs}) is not exact for 
several reasons. On the one hand, $|T_H|^2$ is defined in terms of the 
unprimed momenta $p_{(j)}$, which differ slightly from $p_{(j)}'$. On the 
other hand, the vector $k$ is slightly off shell and has, in general, a 
non-zero transverse component. However, both effects correspond to 
$\Lambda/Q$-corrections, where $Q$ stands generically for the hard scales 
that dominate $T_H$. 

Using Eq.~(\ref{pcs}) the cross section given in Eq.~(\ref{cs2}) can now be 
rewritten as 
\be 
d\sigma=\int dk_-\int\frac{\hat{s}+Q^2}{2\pi q_0}\,(2k_0)^2\,d\hat{\sigma}
(y)\int_{x_\perp}\left|\int\frac{d^2k_\perp}{(2\pi)^2}\,\frac{\mbox{tr}
[\tilde{W}_{x_\perp}(k_\perp'\!\!\!-\!k_\perp)]}{\sqrt{N}\,k^2}\right|^2
\frac{d^3k'}{(2\pi)^3\,2k_0'}\, .\label{cs3}
\ee
The light-cone component $k_-$ is given by $-k_-=yP_-=ym_p$. Note 
that the minus sign in this formula comes from the interpretation of the 
parton with momentum $k$ as an incoming particle in Eq.~(\ref{pcs}). This 
is, in fact, the crucial point of the whole calculation: due to the 
off-shellness of $k$ the corresponding parton can be interpreted as an 
incoming particle in both the process Eq.~(\ref{pp}) and in the soft 
scattering process off the external field. The latter process, where an 
almost on-shell parton with momentum $k$ scatters softly off the external 
field changing its momentum to the on-shell vector $k'$, is most easily 
described in the proton rest frame. By contrast, the natural frame for the 
hard part of the diagram is the Breit frame or a similar frame. In such a 
frame the $k_-$-component is large and negative, so that the above parton 
can be interpreted as an almost on-shell particle with momentum $-k$, 
colliding head-on with the virtual photon. 

It is convenient to use the variables $\xi$ (sometimes called 
$x_{I\!\!P}$), defined by $M^2=(q+\xi P)^2$, and $b$, defined by $b=y/\xi$. 
Note that the latter one is, in general, different from the conventional 
observable $\beta=Q^2/(Q^2+M^2)$. Substituting the variables $y$ and $\xi$ 
for $k_-$ and $k_3'$ the cross section, Eq.~(\ref{cs3}), can be given in 
the form 
\be
\frac{d\sigma}{d\xi}=\int_x^\xi dy\,\hat{\sigma}(y)\left(\frac{df(y,\xi)}
{d\xi}\right)\, ,\label{sx}
\ee
where the diffractive parton distribution for scalars is 
\be
\left(\frac{df(y,\xi)}{d\xi}\right)_{scalar}=\frac{1}{\xi^2}\left(\frac{b}
{1-b}\right)\int\frac{d^2k_\perp'(k_\perp'^2)^2}{(2\pi)^4N}\int_{x_\perp}
\left|\int\frac{d^2k_\perp}{(2\pi)^2}\,\frac{\mbox{tr}[\tilde{W}_{x_\perp}
(k_\perp'\!\!\!-\!k_\perp)]}{k_\perp'^2b+k_\perp^2(1-b)}\right|^2\,.
\label{fx}
\ee
This result is in complete agreement with the concepts developed in 
\cite{bs}. However, the parton distributions introduced there are 
differential in $t=(P'-P)^2$. Since here the proton is modelled by a given 
external field, the present result is automatically inclusive in $t$. 

For an external colour field that is smooth on a soft scale $\Lambda$ and 
confined to a region of approximate size $1/\Lambda$ the function 
$\mbox{tr}[W_{x_\perp}(y_\perp)]$ is also smooth and vanishes at 
$y_\perp=0$ together with its first derivative. From this it can be derived 
that the $k_\perp$- and $k_\perp'$-integrations in Eq.~(\ref{fx}) are 
dominated by the soft scale. This justifies, a posteriori, the softness 
assumption for one of the partons used in the derivation. 

The qualitative result is that the eikonal scattering of this soft parton 
off the proton field determines the diffractive parton distribution. The 
hard part of the photon evolution can be explicitly separated and expressed 
in terms of a standard cross section for photon-parton collisions. 

Having worked out the kinematics in the simple scalar case, it will be 
straightforward to extend the calculation to realistic quarks and gluons in 
the following sections.

\section{Diffractive quark distribution}\label{qua}
The introduction of spinor or vector partons does not affect the 
calculations leading to the generic expression in Eq.~(\ref{sx}). However, 
those parts of the calculation responsible for the specific form of 
Eq.~(\ref{fx}) have to be changed if the soft parton is a spinor or vector 
particle. 

\begin{figure}[ht]
\begin{center}
\vspace*{-.5cm}
\parbox[b]{14cm}{\psfig{width=13cm,file=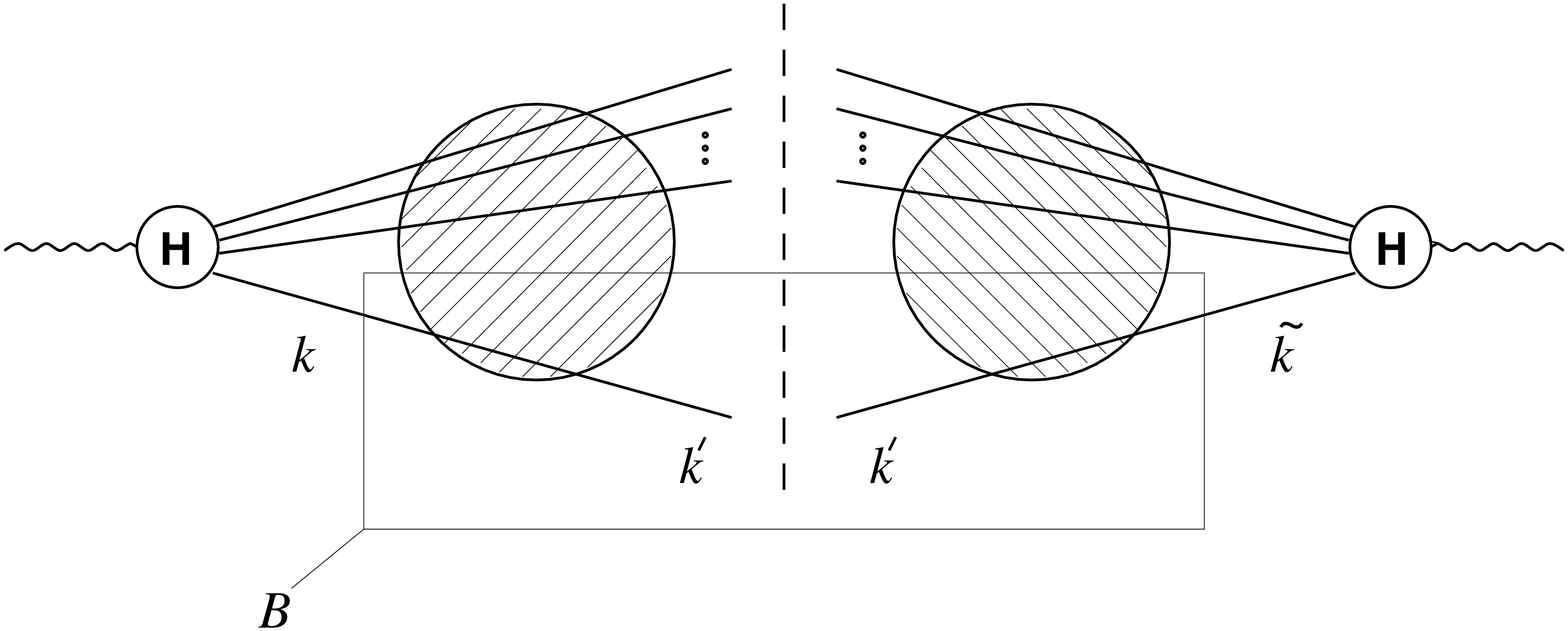}}\\
\end{center}
\refstepcounter{figure}
\label{sp}
{\bf Fig.\ref{sp}} Symbolic representation of the square of the amplitude 
for a hard diffractive process. The box separates the contributions 
associated with the soft parton and responsible for the differences between 
diffractive distributions for scalars, spinors and vector particles.
\end{figure}

The most economic procedure is to first identify the piece of the old 
squared amplitude $|T|^2$ that depends explicitly on the spin of the soft 
parton. This piece, which is essentially just the squared scattering 
amplitude of the soft parton and the external field, is symbolically 
separated in Fig.~\ref{sp}. The two independent integration variables for 
the intermediate momentum of the soft parton are denoted by $k$ and 
$\tilde{k}$ in the amplitude $T$ and its complex conjugate $T^*$ 
respectively. Note also, that $k_+=\tilde{k}_+$ and $k_-=\tilde{k}_-$ at 
leading order. 

It is straightforward to write down the factor $B_{scalar}$ that 
corresponds to the box in Fig.~\ref{sp} for the scalar case. Since the 
off-shell denominators $k^2$ and $\tilde{k}^2$ as well as the two eikonal 
factors and energy $\delta$-functions are present for all spins of the soft 
parton, they are not included into the definition of $B$. All that remains 
are the explicit factors $2k_0$ from the effective vertex, Eq.~(\ref{hev}). 
Therefore, the result for the scalar case reads simply 
\be
B_{scalar}=(2k_0)^2\, .\label{bsc}
\ee

The next step is the calculation of the corresponding expression 
$B_{spinor}$, given by the box in Fig.~\ref{sp} in the case 
where the produced soft parton is a quark. Introducing the factor 
$B_{spinor}/B_{scalar}$ into Eq.~(\ref{fx}) will give the required 
diffractive parton distribution $(df/d\xi)_{spinor}$. 

Observe first, that the analogue of Eq.~(\ref{hev}) for spinors is simply 
\be
V_q(p',p)=2\pi\delta(p_0'-p_0)\,\frac{\gamma_+}{2}\,\tilde{U}(p_\perp'
-p_\perp)\, .\label{hevs}
\ee
The Dirac structure of $V_q$ follows from the fact that in the high energy 
limit only the light-cone component $A_-$ of the gluon field contributes. 
To establish the correctness of the normalization in Eq.~(\ref{hevs}) the 
vertex $V_q$ is used in a scattering amplitude for an on-shell quark. The 
result is
\be
\bar{u}_{s'}(p')V_q(p',p)u_s(p)=2\pi\delta(p_0'-p_0)\,2p_0\,\delta_{s's}\,
\tilde{U}(p_\perp'-p_\perp)\, ,
\ee
in agreement with the  scalar case.

Consider now the Dirac propagator with momentum $k$. The hard part $T_H$ 
requires the interpretation of the quark line as an incoming parton, which 
collides head on with the photon. Therefore, it is convenient to define a 
corresponding on-shell momentum $l$, given by $l_-=-k_-$, $l_\perp=-
k_\perp$ and $l_+=l_\perp^2/l_-$. The propagator can now be written as 
\be
\frac{1}{\ks}=-\frac{\sum_su_s(l)\bar{u}_s(l)}{k^2}-\frac{\gamma_-}{2l_-}
\, .\label{prop}
\ee
This is technically similar to \cite{bdh}, where an on-shell momentum has 
been defined by adjusting the `$-$'-component of $k$. In the present 
treatment, however, the `$+$'-component of $l=-k$ has been adjusted, so 
that a partonic interpretation in the Breit frame becomes possible. It is 
shown in Appendix A, that the second term on the r.h.s. of Eq.~(\ref{prop}) 
can be neglected, since it is suppressed in the Breit frame by the hard 
momentum $l_-=yP_-$. This can be intuitively understood by observing that 
this term represents a correction for the small off-shellness of the quark, 
which is neither important for $T_H$, nor for the soft high energy 
scattering off the external field. 

When the soft part is separated in Fig.~\ref{sp} the spinor $\bar{u}_s(l)=- 
\bar{v}_s(l)$ has to be considered as a part of the hard amplitude $T_H$. 
Therefore the analogue of Eq.~(\ref{bsc}) reads
\be
B_{spinor}=\sum_{s,s'}\,\bar{u}_{s'}(k')\,\frac{\gamma_+}{2}\,u_s(l)\,
\bar{u}_s(\tilde{l})\,\frac{\gamma_+}{2}\,u_{s'}(k')\, ,
\ee
where $\tilde{l}$ is defined analogously to $l$, but using the momentum 
$\tilde{k}$ instead of $k$. The spin summation decouples from the hard part 
if the measurement is sufficiently inclusive for the hard amplitude square 
not to depend on the helicity of the incoming parton. 

The above expression can be evaluated further to give 
\be
B_{spinor}=k_0\sum_s\bar{u}_s(\tilde{l})\gamma_+u_s(l)=k_0\,\frac{4(l_\perp
\tilde{l}_\perp)}{\sqrt{l_-\tilde{l}_-}}\, ,
\ee
where the last equality is most easily obtained using Table II of 
\cite{bl}. Simple kinematics leads to the result 
\be
B_{spinor}=(2k_0)^2\,\frac{2(k_\perp\tilde{k}_\perp)}{k_\perp'^2}\left(
\frac{\xi-y}{y}\right)\, .\label{bsp}
\ee
Comparing this with Eq.~(\ref{bsc}) the diffractive parton distribution is 
straightforwardly obtained from the scalar case, Eq.~(\ref{fx}), 
\be
\left(\frac{df}{d\xi}\right)_{spinor}=\frac{2}{\xi^2}\int\frac{d^2k_\perp'
(k_\perp'^2)}{(2\pi)^4N}\int_{x_\perp}\left|\int\frac{d^2k_\perp}{(2\pi)^2}
\,\frac{k_\perp\mbox{tr}[\tilde{W}_{x_\perp}(k_\perp'\!\!\!-\!k_\perp)]}
{k_\perp'^2b+k_\perp^2(1-b)}\right|^2\, .\label{fxsp}
\ee

The virtual fermion line corresponds to a right-moving quark with momentum 
$k$ in the proton rest frame and to a left-moving antiquark with momentum 
$l$ in the Breit frame. Therefore, the above result has in fact to be 
interpreted as a diffractive antiquark distribution. The diffractive quark 
distribution is identical.

\section{Diffractive gluon distribution}\label{glu}
To obtain the diffractive gluon distribution the procedure of the last 
section has to be repeated for the case of an outgoing soft gluon with 
momentum $k'$ in Fig.~\ref{sp}. Calculating the contribution separated 
by the box will give the required quantity $B_{vector}$, in analogy to 
Eqs.~(\ref{bsc}) and (\ref{bsp}). 

It will prove convenient to introduce two light-like vectors $m$ and $n$, 
such that the only non-zero component of $m$ is $m_-=2$ in the proton rest 
frame, and the only non-zero component of $n$ is $n_+=2$ in the Breit 
frame. Since Breit frame and proton rest frame are connected by a boost 
along the $z$-axis with boost factor $\gamma=Q/(m_px)$, the product of 
these vectors is $(mn)=2\gamma$.

Furthermore, two sets of physical polarization vectors, $e_{(i)}$ and 
$\epsilon_{(i)}$ (with $i=1,2$), are defined by the conditions $ek=\epsilon 
k=0$, $e^2=\epsilon^2=-1$, and $em=\epsilon n=0$. An explicit choice, 
written in light-cone co-ordinates, is
\be
e_{(i)}=\Big(\,0,\,\frac{2(k_\perp\epsilon_{(i)\perp})}{k_+},\,
\epsilon_{(i)\perp}\,\Big)\qquad\mbox{and}\qquad \epsilon_{(i)}=\Big(\,
\frac{2(k_\perp \epsilon_{(i)\perp})}{k_-},\,0,\,\epsilon_{(i)\perp}\,\Big)
\, ,\label{eepsdef}
\ee
where a transverse basis $\epsilon_{(1)\perp}=(1,0)$ and 
$\epsilon_{(2)\perp}=(0,1)$ has been used. Note, that the above equations 
hold in the proton rest frame, in the Breit frame, and in any other frame 
derived by a boost along the $z$-axis.

These definitions give rise to the two following representations for the 
metric tensor: 
\begin{eqnarray}
g^{\mu\nu}&=&\left(\sum_ie_{(i)}^\mu e_{(i)}^\nu+\frac{m^\mu k^\nu}{(mk)}+
\frac{k^\mu m^\nu}{(mk)}-\frac{m^\mu m^\nu}{(mk)^2}k^2\right)\label{gm}
\\
&=&\left(\sum_i\epsilon_{(i)}^\mu\epsilon_{(i)}^\nu\,+\,\frac{n^\mu k^\nu}
{(nk)}\,+\,\frac{k^\mu n^\nu}{(nk)}\,-\,\frac{n^\mu n^\nu}{(nk)^2}k^2\right)
\, .\label{gn}
\end{eqnarray}

The amplitude of the process in Fig.~\ref{sp}, with the lowest parton being 
a gluon in Feynman gauge, is proportional to
\be
A=\epsilon^\mu(k')\,V_g(k',k)_{\mu\nu}T_H^\nu=\epsilon^\mu(k')\,V_g(k',k
)_{\mu\nu}g^{\nu\rho}g_{\rho\sigma}T_H^\sigma\, ,\label{a1}
\ee
where $V_g^{\mu\nu}$ is the effective vertex for the scattering of the 
gluon off the external field. Next, the first and second metric tensor 
appearing in this expression for $A$ are rewritten according to 
Eq.~(\ref{gm}) and Eq.~(\ref{gn}) respectively. It is shown in Appendix B 
that only the first terms from Eqs.~(\ref{gm}) and (\ref{gn}) contribute at 
leading order in $x$ and $\Lambda/Q$. The intuitive reason for this is the 
relatively small virtuality of $k$, which ensures that for both the hard 
amplitude $T_H$ and the soft scattering vertex $V$ only the appropriately 
defined transverse polarizations are important. 

The leading contribution to $|A|^2$, with appropriate polarization 
summation understood, now reads
\be
|A|^2=\!\!\!\!\!\sum_{i,j,i',j',l}\!\left[\Big(\epsilon_{(l)}(k')V_g
e_{(i)}\Big)\Big(e_{(i)}\epsilon_{(j)}\Big)\Big(\epsilon_{(j)}T_H\Big)
\right]\,\left[\Big(\epsilon_{(l)}(k')V_ge_{(i')}\Big)\Big(e_{(i')}
\epsilon_{(j')}\Big)\Big(\epsilon_{(j')}T_H\Big)\right]^*\!\!,
\ee
where the arguments $k$ and $\tilde{k}$ of the polarization vectors in the 
first and second square bracket respectively have been suppressed. 

In the high energy limit the scattering of a transverse gluon off an 
external field is completely analogous to the scattering of a scalar or a 
spinor,
\be
\epsilon_{l'}(p')\,V_g(p',p)\,\epsilon_l(p)=2\pi\delta(p_0'-p_0)\,2p_0\,
\delta_{l'l}\,\tilde{U}_{\cal A}(p_\perp'-p_\perp)\, .\label{eve}
\ee
The only difference comes with the non-Abelian eikonal factor, which is now 
in the adjoint representation \cite{bdh},
\be
U_{\cal A}(x_\perp)={\cal A}(U(x_\perp))\, .
\ee
In analogy to the spinor case, the polarization sum decouples from the hard 
part for sufficiently inclusive measurements, so that the squared amplitude 
is proportional to
\be
|A|^2=|T_H|^2\,\Big(\epsilon_{(l)}(k')V_ge_{(l)}(k)\Big)\Big(\epsilon_{(l)}
(k')V_ge_{(l)}(\tilde{k})\Big)^*\sum_{i,j}\Big(e_{(i)}(k)\epsilon_{(j)}(k)
\Big)\Big(e_{(i)}(\tilde{k})\epsilon_{(j)}(\tilde{k})\Big)\, .\label{aaf}
\ee
Note, that there is no summation over the index $l$. Recall the definition 
of $B$, the soft part of the amplitude square, given at the beginning of 
the last section and illustrated in Fig.~\ref{sp}. The corresponding 
expression in the case of a soft gluon can now be read off from 
Eqs.~(\ref{eve}) and (\ref{aaf}): 
\be
B_{vector}=(2k_0)^2\sum_{i,j}\Big(e_{(i)}(k)\epsilon_{(j)}(k)\Big)\Big(
e_{(i)}(\tilde{k})\epsilon_{(j)}(\tilde{k})\Big)\, .
\ee
This is further evaluated using the explicit formulae in 
Eq.~(\ref{eepsdef}) and the identity 
\be
\sum_{i}\epsilon_{(i)\perp}^a\epsilon_{(i)\perp}^b=\delta^{ab}\qquad
(a,b\in\{1,2\})\, .
\ee
The resulting expression, 
\be
B_{vector}=(2k_0)^2\left(\delta^{ij}+\frac{2k_\perp^ik_\perp^j}{k_\perp'^2}
\left(\frac{1-b}{b}\right)\right)\left(\delta^{ij}+\frac{2\tilde{k}_\perp^i
\tilde{k}_\perp^j}{k_\perp'^2}\left(\frac{1-b}{b}\right)\right)\,,
\ee
is now compared to Eq.~(\ref{bsc}), which gives the diffractive gluon 
distribution 
\be
\left(\frac{df(y,\xi)}{d\xi}\right)_{vector}\!\!\!=\frac{1}{\xi^2}
\left(\frac{b}{1-b}\right)\int\!\frac{d^2k_\perp'(k_\perp'^2)^2}{(2\pi)^4\,
(N^2-1)}\int_{x_\perp}\left|\int\frac{d^2k_\perp}{(2\pi)^2}\,
\frac{\mbox{tr}[\tilde{W}_{x_\perp}^{\cal A}(k_\perp'\!\!\!-\!k_\perp)]\,
t^{ij}}{k_\perp'^2b+k_\perp^2(1-b)}\right|^2\, ,\label{fxg}
\ee
with 
\be
t^{ij}=\delta^{ij}+\frac{2k_\perp^ik_\perp^j}{k_\perp'^2}\left(\frac{1-b}
{b}\right)\, .
\ee
Note, that the factor $N$ appearing in the denominator of Eq.~(\ref{fx}) 
has been replaced by the dimension of the adjoint representation, $N^2-1$.

\section{Simple applications}\label{app}
In this section some results for specific diffractive processes, which can 
be easily obtained within the developed framework, will be discussed.

The lowest order process in leading twist diffraction is the production of 
a colour neutral $q\bar{q}$-pair. This corresponds to Bjorken's aligned jet 
model \cite{bk}, where the relative softness of one of the produced quarks 
gives rise to a large elastic amplitude. It has been shown \cite{bh,bdh} 
that within the eikonal model the aligned jet configuration is indeed 
responsible for the formally leading contribution to diffraction in deep 
inelastic scattering.

In the framework of the present paper the above process is described by 
taking the hard part $T_H$ to be the amplitude for virtual photon-quark 
scattering, $\gamma^*q\to q$. Obviously, in this simplest case neither of 
the quarks has large $p_\perp$. The hard parton is only hard in the sense 
that it carries most of the photon's longitudinal momentum. It can be 
easily checked that this does not invalidate the discussion of 
Sect.~\ref{scal}. 

In the case of one generation of quarks with one unit of electric charge the
well-known partonic cross section reads 
\be
\hat{\sigma}_T(y)^{\gamma^*q\to q}=\frac{\pi e^2}{Q^2}\,\delta(1-y/x)\,.
\ee
The corresponding cross section for longitudinal photons vanishes.

Combining this with the Eqs.~(\ref{sx}) and (\ref{fxsp}) and specifying
to $N=3$ the corresponding contribution to the diffractive structure 
function is easily obtained,
\be
F_2^D(x,Q^2,\xi)^{\gamma^*q\to q}=\frac{2\beta}{3\xi}\int\frac{d^2k_\perp'
(k_\perp'^2)}{(2\pi)^4}\int_{x_\perp}\left|\int\frac{d^2k_\perp}{(2\pi)^2}
\,\frac{k_\perp\mbox{tr}[\tilde{W}_{x_\perp}(k_\perp'\!\!\!-\!k_\perp)]}
{k_\perp'^2\beta+k_\perp^2(1-\beta)}\right|^2\,.\label{f2d}
\ee
Note that in this specific case $y=x$ and $b=\beta=x/\xi$. Adding the 
antiquark contribution and replacing the integration variable $k_\perp'$ by 
$\alpha=k_0'/q_0=k_\perp'^2/M^2$ the result of \cite{bdh} is exactly 
reproduced. 

A simple model for colour neutral quark pair production, based on soft 
colour exchange, had been suggested earlier in \cite{bh1}. The present more 
precise model shows the dominance of the aligned jet configuration and 
replaces the probabilistic colour rotation of \cite{bh1} with the colour 
trace condition in Eq.~(\ref{f2d}). 

Consider next the situation where the diffractive final state contains 
three partons, a $q\bar{q}$-pair and a gluon, two of which have high 
$p_\perp$. The soft parton responsible for leading twist diffraction can be 
either the gluon or one of the quarks (The cross section is the same for 
the $q$- and the $\bar{q}$-case.). Considering the transverse and 
longitudinal photon cross sections separately, four different diffractive 
contributions arise, all of which have been calculated and discussed in 
some detail in \cite{bdh}. In the present framework these contributions 
can be easily obtained using the diffractive quark and gluon distributions 
calculated above together with the partonic cross sections for the 
processes $\gamma^*q\to qg$ and $\gamma^*g\to q\bar{q}$ (see, e.g., 
\cite{field}). 

As an example, the longitudinal photon contribution in the soft quark case 
will be discussed in more detail: the required partonic cross section 
reads 
\be
\frac{d\hat{\sigma}_L(y)^{\gamma^*q\to qg}}{d\hat{t}}=\frac{8e^2\alpha_S}
{3}\,\frac{Q^2(Q^2+\hat{s}+\hat{t})}{(Q^2+\hat{s})^4}\,,
\ee
where $\hat{s}$ and $\hat{t}$ are the usual partonic Mandelstam variables 
and $\hat{s}=(y/x-1)Q^2$.

The corresponding contribution to the longitudinal structure function
can be written as
\begin{eqnarray}
F_L^D(x,Q^2,\xi)^{\gamma^*q\to qg}&=&\frac{\alpha_S}{9\pi^5\xi}
\int_\beta^1db\int d\hat{t}\,\frac{\beta^3(bQ^2+\beta\hat{t})}{b^4Q^4}
\\
&&\times\int d^2k_\perp'(k_\perp'^2)\int_{x_\perp}\left|\int
\frac{d^2k_\perp}{(2\pi)^2}\,\frac{k_\perp\mbox{tr}[\tilde{W}_{x_\perp}
(k_\perp'\!\!\!-\!k_\perp)]}{k_\perp'^2b+k_\perp^2(1-b)}\right|^2\, .
\nonumber
\end{eqnarray}
Adding the antiquark contribution and performing an appropriate change of 
integration variables the corresponding result of \cite{bdh} is again 
exactly reproduced.

The longitudinal photon contribution in the soft gluon case is 
straightforwardly calculated using the diffractive gluon distribution of 
Sect.~\ref{glu}. The result is in agreement with the formula presented 
in \cite{bdh}. Similarly, the cross sections for the remaining two 
transverse photon contributions with a $q\bar{q}g$-final state are found 
to agree with \cite{bdh}. 

Notice finally, that the diffractive parton distributions calculated in the 
present model have to be interpreted as parton distributions at some low 
scale $\sim\Lambda$, i.e., as the input for the QCD evolution equations. 
The simple applications to $q\bar{q}g$-final states discussed in the 
present section are a part of this evolution.

\section{Conclusions}\label{con}
In this paper the calculation of hard diffractive processes in the 
semiclassical model has been organized in a way that shows its 
consistency with the concept of diffractive factorization.

The essentials of the method can already be understood in a model theory of 
scalar partons. Working in the proton rest frame it has been shown that 
high-$p_\perp$ partons from the photon wave function are not sensitive 
kinematically to the colour rotation introduced by the external field. 
Leading twist diffraction appears if a relatively slow parton with small 
$p_\perp$ is found in the photon wave function. The cross section is only 
sensitive to the momentum transfer associated with the impact parameter 
dependence of the eikonal factor of this soft parton. Since the soft parton 
has small negative virtuality, its interpretation as an incoming or 
outgoing particle is ambiguous. Therefore, working e.g. in the Breit frame, 
the hard part of the amplitude can be viewed as describing a head-on 
photon-parton collision. 

As a result, the cross section can be written as a convolution of the hard 
partonic cross section with a function of the non-Abelian eikonal factors, 
that does not involve the hard scale. This function is the diffractive 
parton distribution, calculated in the eikonal model. 

The above factorization of soft and hard parts makes it easy to generalize 
the calculation to a realistic theory with gluons and quarks. In the case 
where the soft parton is a spinor or a vector particle, only the soft part 
of the squared amplitude has to be recalculated. The resulting new 
functions of the eikonal factors represent the diffractive quark and gluon 
distributions. 

In the proton rest frame the above soft parton from the photon wave 
function is still a very fast particle. This parton undergoes a soft 
scattering off the colour field and appears in the diffractive final state 
together with the products of the hard partonic process. Its kinematical 
r\^{o}le is similar to the pomeron remnant appearing in partonic pomeron 
models (compare the discussion in \cite{afs}). Notice however, that no 
pomeron has been introduced in the present calculation. Instead, correlated 
colour singlet parton pairs are found to be a natural consequence of the 
fast proton colour field. 

It appears likely that the concept of diffractive factorization will be 
established as a rigorous consequence of QCD. The present semiclassical 
calculation supplies a QCD-based model for the required diffractive parton 
distributions. The main problem of this model is that once a soft parton 
is found in the photon wave function, further calculations involving 
this parton are not perturbatively justified. In particular, it is not 
clear how large corrections due to additional soft partons in the photon 
wave function and their eikonal scattering will be. 

Nevertheless, it would certainly be interesting to obtain numerical 
predictions for diffractive structure functions from the equations derived 
above and non-perturbative models for the proton colour field. 
\\*[0cm] 

Special thanks go to W.~Buchm\"uller for his continuous support and for 
numerous detailed discussions of different aspects of this investigation. 
I would also like to thank M.~Beneke, S.J.~Brodsky, P.V.~Landshoff, 
M.F.~McDermott, and D.E.~Soper for valuable discussions and comments.

\section*{Appendix A}
The claim that the $\gamma_-$-term in Eq.~(\ref{prop}) can be neglected 
requires some additional considerations. 

The relevant part of the amplitude has the form 
\be
\bar{u}(k')\gamma_+\frac{1}{\ks}T_H=-\bar{u}(k')\gamma_+\left(
\frac{\sum_su_s(l)\bar{u}_s(l)}{k^2}+\frac{\gamma_-}{2l_-}\right)T_H\equiv
-(C_1+C_2)\,.
\ee
Working in the Breit frame, it will be shown that the $\gamma_-$-term $C_2$ 
is suppressed with respect to $C_1$. 

Using the techniques of Sect.~\ref{qua} and assuming $k^2\sim k_\perp^2\sim
k_\perp'^2\sim \Lambda^2$ the first term can be estimated to be 
\be
C_1\sim \frac{\sqrt{\alpha}}{\Lambda}\,\Big[\bar{u}(l)T_H\Big]\,,
\ee
where factors ${\cal O}(1)$ have been suppressed. 

Defining a vector $a$ by the requirements $a_+=1$, $a_-=0$, $a_\perp=0$ in 
the Breit frame, so that $\gamma_-/2=\as$, the $\gamma_-$-term can be 
written as 
\be
C_2=\sum_s\,\Big[\bar{u}(k')\gamma_+u_s(a)\Big]\,\frac{1}{l_-}\,\Big[
\bar{u}_s(a)T_H\Big]\,.
\ee
In the proton rest frame, where both $a_+=q_+/Q$ and $k_+'=\alpha q_+$ are 
large, it is easy to see that $[\bar{u}(k')\gamma_+u(a)]=2\sqrt{a_+k_+'}= 
2q_+\sqrt{\alpha/Q}$ for appropriate helicities. This gives, in the 
Breit frame, $[\bar{u}(k')\gamma_+u(a)]=2Q\sqrt{\alpha/Q}$. Since in this 
frame $l_-$ is a hard momentum $\sim Q$, the following estimate is 
obtained, 
\be
\frac{C_2}{C_1}\,\sim\,\frac{\Lambda}{\sqrt{Q}}\,\frac{[\bar{u}(a)T_H]}
{[\bar{u}(l)T_H]}\,\sim\,\frac{\Lambda}{Q}\,,\label{lq}
\ee
where it has been assumed that no specific cancellation makes 
$\bar{u}(l)T_H$ small, i.e., $[\bar{u}(l)T_H]/[\bar{u}(a)T_H]\sim\sqrt{Q}$. 
Eq.~(\ref{lq}) establishes the required suppression of the $\gamma_-$-term.

\section*{Appendix B}
It has been claimed in Sect.~\ref{glu} that only the first terms of 
Eqs.~(\ref{gm}) and (\ref{gn}), i.e., the transverse polarizations, 
contribute in Eq.~(\ref{a1}) if the metric tensors are rewritten according 
to these formulae. To see this explicitly, consider the expression 
\be
A=\epsilon V_g\left[\sum e\,e+\frac{m\,k}{(mk)}+\frac{k\,m}{(mk)}-
\frac{m\,m}{(mk)^2}(k^2)\right]\,\left[\sum \epsilon\,\epsilon+\frac{n\,k}
{(nk)}+\frac{k\,n}{(nk)}-\frac{n\,n}{(nk)^2}(k^2)\right]T_H\,,\label{aina}
\ee
where the appropriate contractions of vector indices are understood.

Several estimates involving products of $V_g$ and $T_H$ with specific 
polarization vectors will be required.

Note that both $\epsilon$ and $n$ are ${\cal O}(1)$ in the Breit frame. 
For appropriate polarization the amplitude $(\epsilon T_H)$ involves no 
particular cancellation, i.e., it has its leading (formal) power behaviour 
in the dominant scale $Q$. Therefore, $(nT_H)$ is not enhanced with 
respect to $(\epsilon T_H)$,
\be
\frac{(nT_H)}{(\epsilon T_H)}\sim{\cal O}(1)\,.\label{r1}
\ee
By analogy, it can be argued that $(\epsilon V_gm)$ is not enhanced with 
respect to $(\epsilon V_ge)$: since both $e$ and $m$ are ${\cal O}(1)$ in 
the proton rest frame and $(\epsilon V_ge)$ has the leading power behaviour 
for appropriate polarizations $\epsilon$ and $e$, the following estimate 
holds,
\be
\frac{(\epsilon V_gm)}{(\epsilon V_ge)}\sim {\cal O}(1)\,.\label{r2}
\ee

Gauge invariance requires $(kT_H)$ to vanish if $k^2=0$. Since the 
amplitude $T_H$ is dominated by hard momenta ${\cal O}(Q)$, and 
$k^2\sim\Lambda^2\ll Q^2$, this leads to the estimate 
\be
\frac{(kT_H)}{(\epsilon T_H)}\sim\frac{k^2}{Q}\,.\label{r3}
\ee
Analogously, from $(\epsilon V_gk)=0$ at $k^2=0$, the suppression of this 
quantity at small virtualities $k^2$ can be derived,
\be
\frac{(\epsilon V_gk)}{(\epsilon V_ge)}\sim\frac{k^2}{k_+}\,.\label{r4}
\ee
For this estimate it is also important that none of the soft scales 
involved in $V_g$, like $k_\perp^2$ or the gauge field $A$, can appear in 
the denominator to compensate for the dimension of $k^2$. 

All the vector products $nk$, $mk$, $ne$, $m\epsilon$, $mn$, and 
$e\epsilon$ can be calculated explicitly. Using the relations in 
Eqs.~(\ref{r1}), (\ref{r2}), (\ref{r3}), and (\ref{r4}) it is now 
straightforward to show that $(\epsilon V_ge)(e\epsilon)(\epsilon T_H)$ is 
indeed the leading term in Eq.(\ref{aina}). The other terms are suppressed 
by powers of $Q$ or $k_+$.


\begin{thebibliography}{99} 

\bibitem{ex}    H1 collaboration, T. Ahmed et al., Phys. Lett. B348 (1995) 
                681;\\
                ZEUS collaboration, M. Derrick et al., Z. Phys. C68 (1995)
                569

\bibitem{ua8}   UA8 collaboration, A. Brandt et al., Phys. Lett. B297 (1992)
                417

\bibitem{tg}    A.H. Mueller, Nucl. Phys. B335 (1990) 115;\\
                M.G. Ryskin, Sov. J. Nucl. Phys. 52 (1990) 529;\\
                N.N. Nikolaev and B.G. Zakharov, Z. Phys. C53 (1992) 331

\bibitem{is}    G. Ingelman and P.E. Schlein, Phys. Lett. B152 (1985) 256;\\
                A. Donnachie and P.V. Landshoff, Phys. Lett. B191 (1987)
                309; Nucl. Phys. B303 (1988) 634

\bibitem{bs1}   A. Berera and D.E. Soper, Phys. Rev. D50 (1994) 4328

\bibitem{vt}    L. Trentadue and G. Veneziano, Phys. Lett. B323 (1994) 201 

\bibitem{bs}    A. Berera and D.E. Soper, Phys. Rev. D53 (1996) 6162

\bibitem{bh}    W. Buchm\"uller and A. Hebecker, Nucl. Phys. B476 (1996) 203

\bibitem{bdh}   W. Buchm\"uller, M.F. McDermott, and A. Hebecker,
                hep-ph/9607290, to appear in Nucl. Phys. B, Erratum ibid. 

\bibitem{soper} D.E. Soper, talk at the {\it 3rd Workshop on Small-x and
                Diffractive Physics}, Argonne National Laboratory (1996)

\bibitem{nz}    N.N. Nikolaev and B.G. Zakharov, J. Exp. Theor. Phys. 78 
                (1994) 598;\\
                J. Bartels and M. W\"usthoff, J. Phys. G22 (1996) 929;\\
                S.J. Brodsky, P. Hoyer, and L. Magnea, preprint 
                NORDITA-96/68 P,\\ SLAC-PUB-7342, hep-ph/9611278;\\
                M. W\"usthoff, preprint ANL-HEP-PR-97-03, hep-ph/9702201 

\bibitem{css}   J.C. Collins, D.E. Soper, and G. Sterman, Nucl. Phys. B263
                (1986) 37;\\
                O. Nachtmann, Ann. Phys. 209 (1991) 436

\bibitem{bl}    S.J. Brodsky and G.P. Lepage, Phys. Rev. D22 (1980) 2157

\bibitem{bk}    J.D. Bjorken and J.B. Kogut, Phys. Rev. D8 (1973) 1341

\bibitem{bh1}   W. Buchm\"uller and A. Hebecker, Phys. Lett. B355 (1995) 573

\bibitem{field} R.D. Field, {\it Applications of Perturbative QCD}, 
                Addison-Wesley, New York, 1989

\bibitem{afs}   H. Abramowicz, L. Frankfurt and M. Strikman, Proc. of the 
                {\it SLAC Summer Institute 1994} p. 539

\end{thebibliography}
\end{document}